\documentclass[12pt]{article}

\newcommand{\be}{\begin{equation}}
\newcommand{\ee}{\end{equation}}
\newcommand{\bea}{\begin{eqnarray}}
\newcommand{\eea}{\end{eqnarray}}
\newcommand{\beann}{\begin{eqnarray*}}
\newcommand{\eeann}{\end{eqnarray*}}
\newcommand{\nn}{\nonumber}
\newcommand{\ba}{\begin{array}}
\newcommand{\ea}{\end{array}}

\newcommand{\p}{\varphi}
\newcommand{\e}{\epsilon}
\newcommand{\wbar}{\overline}
\newcommand{\diag}{{\rm diag}}

\newcommand{\R}{{\bf R}}

\newcommand{\del}{\partial}

\newcommand{\D}[1]{D$#1$}
\newcommand{\aD}[1]{$\wbar{\mbox{D}#1}$}
\newcommand{\vev}[1]{{\langle #1 \rangle}}

\makeatletter
  
  \@addtoreset{equation}{section}
\makeatother

\begin{document}

%
%
\begin{titlepage}

\setcounter{page}{0}
\renewcommand{\thefootnote}{\fnsymbol{footnote}}

\begin{flushright}
hep-th/0207004\\
KUNS-1793\\
MIT-CTP-3288
\end{flushright}

\vspace{15mm}
\begin{center}
{\Large\bf
Gravitational Approach to Tachyon Matter
}

\vspace{15mm}
{\large
Kazutoshi Ohta$^1$\footnote{e-mail:
kohta@yukawa.kyoto-u.ac.jp}
and
Takashi Yokono$^2$\footnote{e-mail:
yokono@gauge.scphys.kyoto-u.ac.jp}}\\
\vspace{10mm}
{\em $^1$ Center for Theoretical Physics\\
Laboratory for Nuclear Science and Department of Physics\\
Massachusetts Institute of Technology\\
Cambridge, MA 02139, USA}\\
and\\
{\em $^{1,2}$ Department of Physics\\
Kyoto University\\
Kyoto 606-8502, Japan} \\

\end{center}

\vspace{15mm}
\centerline{{\bf{Abstract}}}
\vspace{5mm}

We found a gravity solution of $p{+}1$ dimensional extended object with
 $SO(p)\times SO(9-p)$ symmetry which
 has zero pressure and zero dilaton charge. We expect that this object
 is a residual tachyon dust after tachyon condensation of brane and
 anti-brane system  recently discussed by Sen. 
We also discuss the Hawking temperature and some properties of this
 object.

\end{titlepage}
\newpage

\renewcommand{\thefootnote}{\arabic{footnote}}
\setcounter{footnote}{0}

\section{Introduction}

The studies of dynamics of the tachyon on \D{p}-\aD{p} brane system or
unstable D-branes attract great interests
 for a last few years since they gives the important non-perturbative
 and off-shell 
knowledge of string theory. The analysis using string field theory
(SFT) and effective field theory on the brane is very useful and successful to understand the tachyon
dynamics. We have been naively expecting that the end of an annihilate or
decay of the brane anti-brane or unstable branes goes to nothing by
emitting the graviton while the tachyon is oscillating at the bottom of
the potential. However, more recently, Sen suggested the possibility of
the residual tachyon matter after the tachyon condensation 
where the tachyon is asymptotically rolling
with a constant velocity on the minimum of the runaway potential 
\cite{Sen1,Sen2,Sen3}. This
tachyon matter possess two properties of the absence of plane-wave
solutions around the minimum and the exponential fall off of the
pressure towards zero while the energy is preserved. This kind of
tachyon matter system is discussed from the point of view of boundary
string field theory (BSFT) \cite{ST,Minahan} 
and also could contribute to the inflation or dark
matter problems in cosmology \cite{Gibbons}-\cite{SW}. 

Another approaches to these problems are not so clear and
 insufficient. Especially, if we want to know an annihilation or decay
 process of the unstable D-branes as a bulk gravitational dynamics 
we need a time-dependent decaying gravity
 solutions where the tachyon rolls down to the minimum, but we have not obtained
 completely except for few special examples like S-brane
 solutions \cite{GS,Hashimoto}.
 Alternative view point of the problem is given by
 \cite{BMO,RS,Kim}. They use the general non-BPS $p$-brane solution
 \cite{IM,ZZ} with three
 parameters and argue that these parameters correspond to the charges of
 brane and anti-brane and the vev of the tachyon. We can see the decay
 process as one-parameter deformation, which corresponds to tachyon vev,
 of the gravity solution. Of
 course, for the general value of the tachyon vev open string theory is
 off-shell, but we expect that off-shell open string theory also could
 be probed by scattering of on-shell gravitons and other massless closed
 string through the modified Dirac-Born-Infeld (DBI)
 action. Unfortunately, the three-parameter gravity solution has the
 $ISO(1,p)\times SO(9-p)$ symmetry, which means there is no asymmetry
 between the energy and pressure, and contains no information about vev
 of the tachyon velocity. So this three-parameter
 family solution can not describe the tachyon matter conjectured by Sen.

In this paper, we use more extended solution with four parameters and
the $SO(p)\times SO(9-p)$ symmetry in order to describe the tachyon matter in 
a sense of the gravity of the bulk. We find the energy-momentum tensor on
the brane or boundary. This additional new parameter gives an
asymmetry with respect to the energy and pressure and we expect this
parameter corresponds to the vev of the tachyon velocity. The asymmetry
of the time and space directions in the brane metric allows the pressure
to be zero by tuning the parameters even though the energy is fixed at a
constant value. In addition to the condition that the pressure of the
tachyon matter is zero, we need to set the dilaton charge zero since the
charge is proportional to the world-volume effective DBI action itself
and it vanishes at the minimum of the tachyon potential. So we solve these
conditions and find the gravity solution of the tachyon matter extended
object with the zero pressure and zero dilaton charge. If we start from
the situation of the same number of \D{p}-\aD{p} branes, that is, the
vanishing total RR charge, the solution has only one parameter which
determines the total energy of the system.

This paper is organized as follows. In the next section, we give a brief
review of the rolling tachyon or tachyon matter from the point of view
of the world-volume effective field theory, which was discussed by
Sen. And also we define the energy-momentum tensor and dilaton charge
from the modified DBI action
which are important in the following discussion, and discuss the
relationship between the string frame and Einstein frame quantities. In
Section 3, we calculate the energy-momentum tensor on the brane and dilaton charge
from the bulk gravity side by using the general four-parameter gravity
solution with the $SO(p)\times SO(9-p)$ symmetry. We use these results in
order to determine the tachyon matter solution of the zero pressure and
dilaton charge in Section 4. Finally, we derive the Hawking temperature
of the zero pressure object and discuss on a stability in Section
5. Section 6 is devoted to conclusion and discussion to the future problems.

\section{Rolling Tachyon}

We first consider the effective field theory on the \D{p}-\aD{p} system
or unstable D-branes, which is a modification of the ordinary DBI action
\cite{Garousi,BdRdWEP,Kluson}
\be
S_{BI} = -\tau_p \int d\xi^{p+1}
e^{-\phi}V(T)\sqrt{-\det \hat{A}_{ij}},
\ee
where
\be
\hat{A}_{ij} \equiv \hat{g}_{ij}+\del_i T \del_j T,
\ee
and $T$ is a tachyon field and $V(T)$ is a tachyon potential. We note
that all quantities in the above action like induced metric
$\hat{g}_{ij}$ are defined by the string frame metric of the bulk. We
apply the hatted notation to the string frame quantities in order to
avoid confusing with the Einstein frame ones.
If we
calculate the energy-momentum tensor from the modified DBI action, we obtain
\be
\hat{T}_{ij}
 = -\frac{\tau_p}{4}\frac{e^{-\phi}V(T)}{\sqrt{-\det \hat{g}_{ij}}}
\sqrt{-\det \hat{A}_{ij}}(\hat{A}^{-1})_{ij}.
\ee

If we assume that the tachyon field depends only on time, roughly we have
\bea
\hat{T}_{00} &\simeq& V(T)(\hat{g}_{00}-(\del_0 T)^2)^{-\frac{1}{2}},\\ 
\hat{T}_{ii} &\simeq& -V(T)(\hat{g}_{00}-(\del_0 T)^2)^{\frac{1}{2}}. 
\eea
Therefore if $\hat{g}_{00}-(\del_0 T)^2$ approaches to zero as the same
behaviour as the tachyon potential, $\hat{T}_{00}$ remains constant in
time, but the pressure becomes zero. Actually, if we take the tachyon
potential as $V(T)\propto e^{-\alpha T/2}$ for large $T$, we find a
classical solution of the modified DBI action has the following form for
the large time ($x^0$) \cite{GHY,Sen2}
\be
T=x^0+C e^{-\alpha x^0}+{\cal O}(e^{-2\alpha x^0}),
\ee
and $\hat{T}_{00}$ and $\hat{T}_{ii}$ behave as
\bea
\hat{T}_{00} &\simeq& \frac{1}{\sqrt{2\alpha C}},\\
\hat{T}_{ii} &\simeq& -\sqrt{2\alpha C}e^{-\alpha x^0}.
\eea
So the pressure exponentially dumps to zero with the constant energy as
expected. 

In the following discussion we will treat the above energy-momentum tensor in the
Einstein frame since it is more useful in the gravity
calculation. Rewrite the modified DBI action in the Einstein frame and
calculating the energy-momentum tensor again, we find the same expression
as in the string frame up to a dilaton factor
\be
T_{ij} = -\frac{\tau_p}{4}\frac{e^{-\frac{\phi}{2}}V(T)}{\sqrt{-\det g_{ij}}}
\sqrt{-\det A_{ij}}(A^{-1})_{ij},
\ee
where $A_{ij}$ is now defined by using the Einstein frame induced metric
\be
A_{ij} \equiv e^{\phi/2}g_{ij}+\del_i T \del_j T.
\ee
So the energy-momentum tensor in the string and Einstein frame is
proportional to each other
\be
\hat{T}_{ij} = e^{-\frac{p+3}{4}\phi} T_{ij}.
\ee
Therefore a choice of the frames does not affect the discussion of the tachyon
matter. If we can find the pressureless object in the Einstein frame, we
expect zero pressure in the string frame too.

We also would like the another quantity which is a variation with
respect to the dilaton field
\be
\frac{\delta S_{BI}}{\delta \phi}
=\tau_p e^{-\phi} V(T) \sqrt{\det \hat{A}_{ij}}.
\ee
This is the modified DBI Lagrangian itself. So this quantity has to
vanish at the minimum of the tachyon potential since there is no
world-volume effective theory after the condensation. We note that
this variation depends on the frame. If we use the Einstein frame, we
can not conclude that the variation vanish at minimum. So we have to use
the string frame on this quantity.
We refer this quantity as a dilaton charge in the following gravity
discussions.

\section{Four-Parameter Brane Solution and Energy\\-momentum Tensor}

\subsection{Gravity solution}

The complete solution for brane like extended objects was found by
\cite{IM,ZZ}. They solve only the equation of motion for a coupled
system of gravity, dilation and RR anti-symmetric tensor fields in any
dimensions. The ansatz for a $p+1$ dimensional extend object has
$SO(p)\times SO(9-p)$ symmetry, where $SO(p)$ and $SO(9-p)$ are the
symmetry of the space
directions on the brane-like object  and the $9-p$ dimensional transverse
directions. This solution includes the well-known BPS \D{p}-brane, where
the symmetry on the brane enhances to $ISO(1,p)$ Poincar\'e
invariance. For a general non-extremal or non-BPS branes, the Poincar\'e
symmetry breaks to $SO(p)$.

The $SO(p)\times SO(9-p)$ symmetric solution includes four
parameters. If we now use the notations of these parameters in
\cite{BMO}, the metric of four-parameter solution is expressed in the
Einstein frame as
\bea
ds^2 &=& e^{2A(r)}\left(-f(r)dt^2+dx_{m}dx^{m}\right)
+e^{2B(r)}(dr^2+r^2d\Omega_{8-p}^2),\nn\\
\phi&=&\phi(r),\label{4para solution}\\
C^{(p+1)}&=&e^{\Lambda(r)}dx^0\wedge \ldots \wedge dx^p,\nn
\eea
where
\bea
f(r)&=&e^{-c_3h(r)},\label{c3}\\
A(r)&=&\frac{7-p}{32}
\left(\frac{3-p}{2}c_1+\left(1+\frac{(3-p)^2}{8(7-p)}\right)c_3\right)h(r)\nn\\
&&-\frac{7-p}{16}\ln\left[\cosh(kh(r))-c_2\sinh(kh(r))\right],\\
B(r) &=&\frac{1}{7-p}\ln\left[f_-(r)f_+(r)\right]
+\frac{p-3}{64}\left((p+1)c_1-\frac{3-p}{4}c_3\right)h(r)\nn\\
&&+\frac{p+1}{16}\ln\left[\cosh(kh(r))-c_2\sinh(kh(r))\right]\\
\phi(r)&=&\frac{7-p}{16}\left((p+1)c_1-\frac{3-p}{4}c_3\right)h(r)\nn\\
&&+\frac{3-p}{4}\ln\left[\cosh(kh(r))-c_2\sinh(kh(r))\right],\\
e^{\Lambda(r)}&=&-\frac{\sqrt{c_2^2-1}\sinh(kh(r))}{\cosh(kh(r))-c_2\sinh(kh(r))},
\eea
and
\bea
f_{\pm}(r)&=&1\pm\left(\frac{r_0}{r}\right)^{7-p},\label{f(r)}\\
h(r)&=&\ln\left[\frac{f_-(r)}{f_+(r)}\right],\label{h(r)}\\
k^2&=&\frac{2(8-p)}{7-p}-c_1^2
+\frac{1}{4}\left(\frac{3-p}{2}c_1+\frac{7-p}{8}c_3\right)^2
-\frac{7}{16}{c_3}^2.
\eea
As we can see in the above solution, the asymmetry between the time and
space directions on the brane is brought by the parameter $c_3$ in
$f(r)$ of eq.~(\ref{c3}) only. So
if we set $c_3=0$, the $ISO(1,p)$ symmetry of the world-volume is
restored.

The detailed analysis for the three-parameter solution with
$c_3=0$ is
discussed in \cite{BMO}. The solution is invariant under the $Z_2$
transformations of these three parameters $(r_0,c_1,c_2)$ and we can
fix a physical range of parameters as $r_0,\ c_1,\ k \in \R_+$ and
$c_2\in (-\infty,-1)\cup(1,\infty)$. 

The total RR charge vanishes if $|c_2|=1$, so we expect $|c_2|=1$
corresponds to the situation that the number of \D{p} ($N$) and \aD{p}
($\wbar{N}$) branes
is the same ($N=\wbar{N}$). Moreover, in this neutral charged case, the physically
relevant choice is $c_2=1$ for $p>3$ and $c_2=-1$ for $p<3$ (for $p=3$
the choices of sign are physically equivalent).

On the other hand, the BPS \D{p}-brane ($N\neq 0,\ \wbar{N}=0$) 
is appeared in the limit of
$c_2\rightarrow \infty$ for $p\geq3$ or $c_2\rightarrow -\infty$ for
$p<3$. If we consider the following scaling relations for $p>3$
\bea
r_0^{7-p} &\rightarrow& \e^{\frac{1}{2}}r_0^{7-p},\nn\\
c_1 &\rightarrow& c_m-\e\frac{8k^2}{(p+1)(7-p)c_m},\nn\\
c_2 &\rightarrow& \frac{c_2}{\e},\nn
\eea
where $c_m=\left(\frac{32(8-p)}{(p+1)(7-p)^2}\right)^{\frac{1}{2}}$ and
\be
\mu_0 = 2 c_2 k r_0^{7-p}
\ee
is fixed, then we finally obtain the metric of the BPS \D{p}-brane \cite{HS}
\bea
ds^2 &=& H^{\frac{p-7}{8}}dx_\mu dx^\mu 
+ H^{\frac{p+1}{8}}(dr^2+r^2d\Omega_{8-p}^2),\nn\\
e^{\phi}&=&H^{\frac{3-p}{4}},\nn\\
C_{01\ldots p}^{(p+1)} &=& -\frac{1}{2}(H^{-1}-1),\nn\\
H&=&1+\frac{\mu_0}{r^{7-p}}.\nn
\eea

The meaning of the parameter $c_1$ is also discussed in \cite{BMO}. The
authors conclude that $c_1=0$ corresponds to $\vev{T}=0$
where $\vev{T}$ is a
vev of the tachyon field and $c_1$ mixed with $r_0$ has a non-trivial
relation to $\vev{T}$. However, it is too difficult to find exact
maps between $c_1,\ r_0$ and $\vev{T}$.

Now let us turn on the parameter $c_3$. The parameter $c_3$ breaks the
world-volume symmetry to $SO(p)$ and the solution (\ref{4para solution})
with four independent parameters $(r_0,c_1,c_2,c_3)$ includes the black
$p$-brane solutions of \cite{HS} at
$(c_1,c_3)=\left(\frac{3-p}{2(7-p)},-2\right)$ and the Schwarzschild
solution at $(c_1,c_2,c_3)=\left(\frac{3-p}{2(7-p)},\pm
1,-2\right)$. Introduction of $c_3$ produce the asymmetry between time
and space component of the energy-momentum tensor on the brane as we will see in the
following section. And also as we have mentioned in the previous section,
the velocity of tachyon $\del_0 T$ gives the different behaviour of the
energy and pressure. Therefore, although we can not determine the exact
relation, we expect that there is a relation
between the parameter $c_3$ and a vev of the tachyon velocity
$\vev{\del_0 T}$. We will show in the following that a suitable choice of the parameters
$c_1$ and $c_3$ makes a tachyon matter gravity solution with a positive
energy and zero pressure. 

\subsection{Brane system and energy-momentum tensor}

We begin with the bosonic part of Type IIA/B supergravity action in ten
dimensions coupled with a boundary brane action
\bea
S &=& S_G + S_B,
\label{action}\\
S_G &=& \frac{1}{16\pi G_{10}}
\int d^{10}x \sqrt{-g}\left\{R
-\frac{1}{2}\del_\mu\phi\del^\mu\phi
-\frac{1}{2(p+2)!}e^{\frac{3-p}{2}\phi}F_{(p+2)}^2
\right\},\\
S_B&=&\int d^{10}x \delta^{(9-p)}(r){\cal L}_B,
\eea
where $S_G$ represents a bulk supergravity action of the graviton,
dilaton and anti-symmetric RR fields, and $S_B$ is a boundary term which
has a delta function distribution on the brane.
The Lagrangian ${\cal L}_B$ in the action $S_B$ is the DBI type
Lagrangian of BPS branes
\be
{\cal L}_B = -\tau_p e^{-\phi}
\sqrt{-\det\left[e^{\frac{\phi}{2}}g_{ij}+2\pi\alpha' F_{ij}\right]}
\ee
 with tension $\tau_p$, or the modified DBI Lagrangian of non-BPS branes,
 which we have used in Section 2
\be
{\cal L}_B = -\tau_p e^{-\phi}
V(T)\sqrt{-\det\left[e^{\frac{\phi}{2}}g_{ij}+2\pi\alpha' (F_{ij}+\del_i T\del_j T)\right]},
\ee
which contains the tachyon field $T$ and tachyon
potential $V(T)$ and we denote these actions in the Einstein frame.

The Einstein equations of motion derived from (\ref{action}) is
\bea
R_{\mu\nu}-\frac{1}{2}g_{\mu\nu}R &=& 8\pi G_{10}T^{total}_{\mu\nu}\nn\\ 
&=& 8\pi G_{10}\left(T^{bulk}_{\mu\nu} +
\delta^{(9-p)}(r)T_{\mu\nu}
\right),\label{em}
\eea
where $T^{bulk}_{\mu\nu}$ is an
energy-momentum tensor contributing from the dilaton field and RR
(p+2)-form and $T_{\mu\nu}$ is an energy-momentum tensor localized
on the branes. The solution (\ref{4para solution}), of course, simply 
satisfies
the equation motion (\ref{em}) on the all region except for at $r=0$,
but our present interested part of the energy-momentum tensor is the
localized  one on the brane
which is the coefficient of the delta-function of $r$.

If we want to extract the delta functional distribution part of the
energy-momentum tensor from the general brane metric 
(\ref{4para solution}),
we have to pay some attention to obtain it.

To see this, we first expand the metric around the asymptotically flat
region as like as
\be
g_{\mu\nu} = \eta_{\mu\nu} + h_{\mu\nu},
\ee
where $\eta_{\mu\nu} = \diag(-1, +1 ,+1 \cdots)$. If we now define
\be
\p_{\mu\nu} \equiv h_{\mu\nu}-\frac{1}{2}\eta_{\mu\nu}h,
\ee
where $h={h^{\alpha}}_\alpha$ and use the fact that the Einstein
 frame metric satisfies the following 
harmonic gauge
\be
\del_\mu \p^{\mu\nu} = 0,
\ee
then we obtain the following linear approximation of the Einstein
equation up to the order one of $h_{\mu\nu}$
\be
\frac{1}{2}\nabla^2\p_{\mu\nu} = -8\pi G_{10} T^{total}_{\mu\nu}.
\label{linear}
\ee
Following eq.~(\ref{linear}), the delta function of $r$ in the
energy-momentum tensor comes from the singular part of the order
$1/r^{7-p}$ in $\p_{\mu\nu}$ by action of the Laplacian in the
transverse space. 

Expanding the metric of the solution (\ref{4para solution}) as
\bea
g_{00} &=& -1 
+\alpha
\left(\frac{r_0}{r}\right)^{7-p} + \cdots,\\
g_{ii} &=& 1
+ \beta 
\left(\frac{r_0}{r}\right)^{7-p} + \cdots,\\
g_{MM} &=& 1
+ \gamma
\left(\frac{r_0}{r}\right)^{7-p} + \cdots,
\eea
where $i=1,\cdots,p$ and $M=p+1,\cdots,9$ represent the indices of the
longitudinal and transverse space, respectively, and here we define
\bea
\alpha &=& \frac{1}{64}\left(16 (7{-}p)k c_2 -4(p{-}3)(7{-}p)c_1
 + (p^2{-}14p{-}63)c_3\right),\\
\beta &=& \frac{1}{64}\left(- 16 (7{-}p)k c_2 +4(p{-}3)(7{-}p)c_1
 - (p^2{-}14p{+}65)c_3\right),\\
\gamma &=& \frac{1}{64}\left(16 (p{+}1)k c_2 -4(p{+}1)(p{-}3)c_1
  - (p{-}3)^2c_3\right).
\eea

Finally we have an expression for the energy momentum tensor on the brane
in the Einstein frame.
\bea
T_{00} &=& \frac{1}{2}\left(
\alpha+p\beta+(9{-}p)\gamma\right)
\frac{(7-p)V_p\omega_{8-p}r_0^{7-p}}{16\pi
G_{10}},\nn\\
&=& \left(16c_2k-4(p{-}3)c_1-(9{+}p)c_3\right)
\frac{(7-p)V_p\omega_{8-p}r_0^{7-p}}{16\pi
G_{10}},\\
T_{ii} &=& \frac{1}{2}\left(
\alpha-(p{-}2)\beta-(9{-}p)\gamma\right)
\frac{(7-p)V_p\omega_{8-p}r_0^{7-p}}{16\pi
G_{10}},\nn\\
&=& -\left(16c_2k-4(p{-}3)c_1+(7{-}p)c_3\right)
\frac{(7-p)V_p\omega_{8-p}r_0^{7-p}}{16\pi
G_{10}},\label{pressure}\\
T_{MM} &=& \frac{1}{2}\left(
\alpha-p\beta-(7{-}p)\gamma\right)
\frac{(7-p)V_p\omega_{8-p}r_0^{7-p}}{16\pi
G_{10}},\nn\\
&=& 0,
\eea
where we use the relation
\be
\nabla^2\left(\frac{1}{r^{7-p}}\right)
= -(7-p)V_p\omega_{8-p}\delta^{(9-p)}(r),
\ee
where $V_p$ is a normalized volume of the brane and $\omega_{8-p}$ is a
volume of $8-p$ dimensional unit sphere.

$T_{MM}=0$ means that the object is not extended in transverse space
and all components of the energy-momentum tensor are localized on the
brane at $r=0$. $T_{00}$ is well known as the Arnowitt-Deser-Misner (ADM)
 mass $M_{\mbox{\tiny
ADM}}$, which agrees with other derivations of \cite{MP,Argurio}. 
As we have noted before, if we set $c_3=0$ the
world-volume symmetry is restored to $ISO(1,p)$ and we have a
restricted relation $T_{00}=-T_{ii}$. So we can not reach at the situation
of the tachyon matter without the parameter $c_3$.

\subsection{Dilaton charge}

Following the discussion in Section 2, the variation of the unstable
\D{p}-brane action with respect to the dilation field must vanish at the
minimum of the tachyon potential. This means that the source term of the
bulk dilation filed also vanishes. Since this argument is frame dependent, we
derive the string frame equation of motion of the dilaton from the action (\ref{action})
\be
\sqrt{-\hat{g}}e^{-2\phi}\left(\hat{R}+4\del_\mu\phi\del^\mu\phi\right)
+4\del_\mu\left(\sqrt{-\hat{g}}e^{-2\phi}\hat{g}^{\mu\nu}\del_\nu\phi\right)
= -8\pi G_{10} Q_D \delta^{(9-p)}(r).
\ee
The contribution to the delta function comes from the total derivative
term in $\hat{R}$ and Laplacian of $\phi$. If we again use the weak field
approximation around the asymptotically flat region, we have the
following linearized Laplace equation
\be
\nabla^2 \left(\hat{h}_{MM}-\hat{h}+4\phi\right)
=-8 \pi G_{10} Q_D \delta^{(9-p)}(r).
\ee
Picking up the delta function pole from the order of $1/r^{7-p}$, we
obtain
\be
Q_D = \left(16 c_2 k-4(p{+}1)c_1-(p{-}3)c_3\right)
 \frac{(7-p)V_p\omega_{8-p}r_0^{7-p}}{128\pi G_{10}}.
\label{dilaton charge}
\ee
$Q_D=0$ is a necessary condition at the minimum of the tachyon potential.

From the world-volume effective theory analysis, we can see that the space
components of the energy-momentum tensor $\hat{T}_{ii}$ and the dilation
charge are proportional to each other up to some overall factor, but
(\ref{pressure}) and (\ref{dilaton charge}) in our analysis from the gravity
side are obviously not proportional.
However, if we
use the effective filed theory from BSFT \cite{ST,Minahan}, the pressure
and dilation charge are simply not proportional to each other. From our
standpoint, we can not determine the explicit coupling of the bulk
fields to world-volume theory. So we do not straightaway think this fact
is contradiction.

\section{Tachyon Matter}

Now we solve the condition for the tachyon matter, namely $T_{ii}=0$ and
$Q_D=0$ while $T_{00}\neq 0$. In the following arguments, we assume $3<p<7$
and set $c_2=1$ since we would like to fix a physical region of the
parameters and consider the case of the vanishing total RR charge
($N=\wbar{N}$). From the eqs.~(\ref{pressure}) and
(\ref{dilaton charge}), we find
\bea
(c_1,c_3) &=& 
\left(
\frac{1}{2}\sqrt{\frac{8-p}{7-p}},-2\sqrt{\frac{8-p}{7-p}}
\right)\label{c1c3}\\
k &=& \frac{1}{2}\sqrt{\frac{8-p}{7-p}}\label{k},
\eea
where there is an alternative choice of sign in $c_1$, $c_3$ and $k$ as a
solution,
 but we have
fixed by using the physical condition of $T_{00}\geq 0$. Therefore, in
the above parameter choice, the energy or ADM mass is positive definite
\be
M_{\mbox{\tiny ADM}}
=\frac{2\sqrt{(8-p)(7-p)}V_p\omega_{8-p}r_0^{7-p}}{\pi G_{10}},
\ee
which includes only one parameter of $r_0\geq 0$. We also note that all
parameters are in the physical region if $p<7$.

Using eqs.~(\ref{c1c3}) and (\ref{k}), we find the relation of 
$(c_1,c_3)=(k,-4k)$ and substituting into the metric (\ref{4para
solution}), we have the metric of the tachyon matter in the Einstein frame
\bea
ds^2&=&e^{-\frac{1}{2}kh(r)}
\left[
-e^{4kh(r)}dt^2+dx_m dx^m+\left(f_-(r)f_+(r)\right)^{\frac{2}{7-p}}
\left(dr^2+r^2 d\Omega_{8-p}^2\right)
\right],\label{Einstein}\\
\phi(r)&=&kh(r)\label{dila},
\eea
where $h(r)$ and $f_\pm(r)$ are also given by (\ref{h(r)}) and
(\ref{f(r)}). This metric has the $SO(p)\times SO(9-p)$ symmetry and is
static. This static gravity solution could describe the tachyon matter object
which reaches a constant tachyon vev $\vev{T}$ and velocity
 $\vev{\partial_0 T}$
after the rolling tachyon condensation.

If we now rewrite (\ref{Einstein}) to the string frame metric by using relation
$\hat{g}_{\mu\nu}=e^{\frac{\phi}{2}}g_{\mu\nu}$, we obtain
\be
d\hat{s}^2=
-\left(\frac{f_-(r)}{f_+(r)}\right)^{2\sqrt{\frac{8-p}{7-p}}}dt^2
+dx_m dx^m
+\left(f_-(r)f_+(r)\right)^{\frac{2}{7-p}}
\left(dr^2+r^2 d\Omega_{8-p}^2\right),
\label{string}
\ee
and the dilation field is again given by (\ref{dila}).

This solution is controlled only by the parameter $r_0$. If the total
energy of the system decreases due to the emission of the graviton or
evaporation, $r_0$ also must decrease. In the limit of $r_0\rightarrow 0$,
we have $f_\pm(r)\rightarrow 1$. 
So, the metric (\ref{Einstein}) or (\ref{string}) become a flat
10-dimensional space in the limit of $M_{\mbox{\tiny ADM}}\rightarrow 0$
as expected.

\section{Hawking temperature}

To discuss a quantum stability of the object found in Section 4, we need
the semi-classical thermodynamics of the above gravitational system,
namely the Hawking temperature of the brane-like object.
The Hawking temperature is determined by avoiding conical singularity on
the plane of the Euclidean time and the radial coordinate. The some useful
expressions of the Hawking temperature and entropy for the general
 black $p$-brane solution are derived in \cite{Argurio}.

In our case of $c_2=1$, if we define 
$\e^{7-p}\equiv (r^{7-p}-r_0^{7-p})/r_0^{7-p}$,
the Hawking temperature is given by the surface gravity at the horizon
($r\rightarrow r_0$ or $\e\rightarrow 0$)
\bea
T_H &=& \left.\frac{1}{2\pi}
\left[
\frac{\frac{d}{dr}\left(f(r)^{1/2}e^{A(r)})\right)}{e^{B(r)}}
\right]\right|_{r\rightarrow r_0},\nn\\
&=& \left.
\frac{\xi}{2\pi r_0}\e^{\lambda}\right|_{\e\rightarrow 0},
\eea
where
\bea
\xi &=& \frac{7{-}p}{128}
\left(16(7{-}p)k -4(p{-}3)(7{-}p)c_1+(p^2{-}14p{-}63)c_3\right),\\
\lambda &=& -(8{-}p)
+\frac{7{-}p}{32}\left(
16k - 4(p{-}3)c_1 -(p{+}9)c_3
\right).
\eea
This result means that if $\lambda>0$ then $T_H=0$ and if $\lambda=0$ then
$T_H$ is finite, but if $\lambda<0$ then the Hawking temperature
diverges as long as $\xi$ is a positive constant. 

Now substituting our choice of the parameters in the pressureless
object (\ref{c1c3}) and (\ref{k}), we find
\bea
\xi&=& \frac{7}{4}\sqrt{(8-p)(7-p)},\\
\lambda &=& -(8-p)+\sqrt{(8-p)(7-p)},
\eea
and $\lambda$ is always negative for $p<7$. This signals that the
Hawking temperature of the object we found is infinite and using the
Hawking radiation argument the object is highly unstable and may
evaporate and disperse.

However, this conclusion of the semi-classical analysis is presumably
too naive since we do not use any information about non-interacting closed
string pressureless gas 
localized on the tachyon matter at the horizon. We need further
discussions for the stability of this object.

\section{Conclusion and Discussion}

In this paper, we found the metric of the $p+1$ dimensional extended
object without the pressure and dilation charge. However, there is no
conclusive evidence that our gravity solution describes the tachyon
matter discussed by Sen since we do not have sufficient knowledge of
the relationship between the world-volume open string tachyon and
parameters in the closed
string bulk gravity solution.

We also do not know the explicit relation between the vevs in the
world-volume effective theory $(\vev{T},\vev{\del_0 T})$ and the
parameters in the gravity solution $(c1,c3)$. Information about the
correspondence gives justification on the gravity description of the
tachyon matter.

To understand more deeply the nature of the tachyon matter,
 we have to develop closed string theory on the
background (\ref{string}). String theory has a upper limit of the
temperature which is known as the Hagedorn temperature. So our
semi-classical analysis on the Hawking temperature may fail at the
Hagedorn temperature of the closed string on the brane. We hope that a string
dynamics at high temperature gives us reason for zero pressure and true
properties of the tachyon matter.

\section*{Acknowledgements}

We would like to thank S.~Kawamoto for useful discussions. 
KO would like to thank A.~Fayyazuddin and A.~Sen for very useful discussions
and comments. KO would also like to thank B.~Zwiebach for a lucid
lecture on tachyon matter. TY would like to thank M.~Fukuma,
K.~Hashimoto, K.~Hotta for useful discussions.
The work of KO and TY is supported in part by Japan Society for the
Promotion of Science Research Fellowship (\#02809 and
\#02845). KO is also supported in part by funds provided by the
U.S. Department of Energy (D.O.E.) under cooperative research agreement
\#DF-FC02-94ER40818.

%
%
\newpage

\newcommand{\NP}[3]{Nucl.~Phys.~{\bf #1} (#2) #3}
\newcommand{\PL}[3]{Phys.~Lett.~{\bf #1} (#2) #3}
\newcommand{\PR}[3]{Phys.~Rev.~{\bf #1} (#2) #3}
\newcommand{\PRL}[3]{Phys.~Rev.~Lett.~{\bf #1} (#2) #3}
\newcommand{\AP}[3]{Ann.~Phys.~{\bf #1} (#2) #3}
\newcommand{\CMP}[3]{Comm.~Math.~Phys.~{\bf #1} (#2) #3}
\newcommand{\JHEP}[3]{JHEP {\bf #1} (#2) #3}
\newcommand{\JMP}[3]{J.~Math.~Phys.~{\bf #1} (#2) #3}
\newcommand{\hepth}[1]{{\tt hep-th/#1}}

\newcommand{\lit}[3]{#1, ``#2'', #3.}


\begin{thebibliography}{99}


\bibitem{Sen1}
    \lit{A.~Sen}
        {Rolling Tachyon}
        {\JHEP{0204}{2002}{048}, \hepth{0203211}}

\bibitem{Sen2}
    \lit{A.~Sen}
        {Tachyon Matter}
        {\hepth{0203265}}

\bibitem{Sen3}
    \lit{A.~Sen}
        {Field Theory of Tachyon Matter}
        {\hepth{0204143}}

\bibitem{ST}
    \lit{S.~Sugimoto, S.~Terashima}
        {Tachyon Matter in Boundary String Field Theory}
        {\hepth{0205085}}

\bibitem{Minahan}
    \lit{J.~Minahan}
        {Rolling the tachyon in super BSFT}
        {\hepth{0205098}}

\bibitem{Gibbons}
    \lit{G.~W.~Gibbons}
        {Cosmological Evolution of the Rolling Tachyon}
        {\PL{B537}{2002}{1}, \hepth{0204008}}

\bibitem{FT}
    \lit{M.~Fairbairn, M.~H.~G.~Tytgat}
        {Inflation from a Tachyon Fluid?}
        {\hepth{0204070}}

\bibitem{Mukohyama}
    \lit{S.~Mukohyama}
        {Brane cosmology driven by the rolling tachyon}
        {\hepth{0204084}}

\bibitem{Feinstein}
    \lit{A.~Feinstein}
        {Power-Law Inflation from the Rolling Tachyon}
        {\hepth{0204140}}

\bibitem{Padmanabhan}
    \lit{T.~Padmanabhan}
        {Accelerated expansion of the universe driven by tachyonic matter}
        {\hepth{0204150}}

\bibitem{FKS}
    \lit{A.~Frolov, L.~Kofman, A.~Starobinsky}
        {Prospects and Problems of Tachyon Matter Cosmology}
        {\hepth{0204187}}

\bibitem{CGJP}
    \lit{D.~Choudhury, D.~Ghoshal, D.~P.~Jatkar, S.~Panda}
        {On the Cosmological Relevance of the Tachyon}
        {\hepth{0204204}}

\bibitem{SW}
    \lit{G.~Shiu, I.~Wasserman}
        {Cosmological Constraints on Tachyon Matter}
        {\hepth{0205003}}

\bibitem{GS}
    \lit{M.~Gutperle, A.~Strominger}
        {Spacelike Branes}
        {\JHEP{0204}{2002}{018}, \hepth{0202210}}

\bibitem{Hashimoto}
    \lit{K.~Hashimoto}
        {Dynamical Decay of Brane-Antibrane and Dielectric Brane}
        {\hepth{0204203}}

\bibitem{BMO}
    \lit{P.~Brax, G.~Mandal, Y.~Oz}
        {Supergravity Description of Non-BPS Branes}
        {\PR{D63}{2001}{064008}, \hepth{0005242}}

\bibitem{RS}
    \lit{R.~Rabadan, J.~Simon}
        {M-theory lift of brane-antibrane systems and localised closed
         string tachyons}
        {\JHEP{0205}{2002}{045}, \hepth{0203243}}

\bibitem{Kim}
    \lit{H.~Kim}
        {Supergravity Approach to Tachyon Potential in Brane-Antibrane Systems}
        {\hepth{0204191}}

\bibitem{IM}
    \lit{V.~D.~Ivashchuk, V.~N.~Melnikov}
        {Multidimensional Classical and Quantum Cosmology with
	Intersecting p-branes}
        {\JMP{39}{1998}{2866-2888}, \hepth{9708157}}

\bibitem{ZZ}
    \lit{B.~Zhou, C-J.~Zhu}
        {The Complete Black Brane Solutions in D-dimensional Coupled
 Gravity System}
        {\hepth{9905146}}

\bibitem{Garousi}
    \lit{M.~R.~Garousi}
        {Tachyon couplings on non-BPS D-branes and Dirac-Born-Infeld action}
        {\NP{B584}{2000}{284-299}, \hepth{0003122}}

\bibitem{BdRdWEP}
    \lit{E.~A.~Bergshoeff, M.~de Roo, T.~C.~de Wit, E.~Eyras, S.~Panda}
        {T-duality and Actions for Non-BPS D-branes}
        {\JHEP{0005}{2000}{009}, \hepth{0003221}}

\bibitem{Kluson}
    \lit{J.~Kluson}
        {Proposal for non-BPS D-brane action}
        {\PR{D62}{2000}{126003}, \hepth{0004106}}

\bibitem{GHY}
    \lit{G.~Gibbons, K.~Hori, P.~Yi}
        {String Fluid from Unstable D-branes}
        {\hepth{0009061}}

\bibitem{DS}
    \lit{M.~J.~Duff, K.~S.~Stelle}
        {Multi-membrane solutions of D=11 supergravity}
        {\PL{B253}{1991}{113}}

\bibitem{HS}
    \lit{G.~T.~Horowitz, A.~Strominger}
        {Black Strings and p-Branes}
        {\NP{B360}{1991}{197}}

\bibitem{MP}
    \lit{R.~C.~Myers, M.~J.~Perry}
        {Black Holes in Higher Dimensional Space-Time}
        {\AP{172}{1986}{304}}

\bibitem{Argurio}
    \lit{R.~Argurio}
        {Brane Physics in M-theory}
        {\hepth{9807171}}








\end{thebibliography}
\end{document}